# Deformed Microcavity for Far Field Biosensing

**Qinghai Song, Young L Kim**

*Weldon School of Biomedical Engineering, Purdue University, West Lafayette, IN, USA, 47907*

**Abstract:**

Here we demonstrate a new concept for designing an ultra-sensitive deformed cavity biosensor. Owning to the breaking of rotational symmetry, the field distribution is not uniform along the cavity boundary and results in the dependence of spectra shift and mode splitting on the position of a scatter. In this case, the deformed cavity sensor can be extremely sensitive to the location of particle binding on the cavity boundary. Moreover, the directional emission from the deformed microcavity provides a possibility to detect a single particle or molecule in the far field.

Optical microcavities have been the object of intensive investigation for several important applications including nanophotonics circuits, integrated optical chips, and optical communications [1, 2]. Moreover, microcavities have recently been proposed as a powerful method to achieve sensitive label-free biosensors because of the ultrahigh Q factors of whispering gallery modes (WGM) [3]. It was experimentally demonstrated that the highest Q factor in microspheres can be as high as $8\times10^9$ [4], which in turn provides the ultrahigh sensitivity for detecting single molecules using a resonant wavelength shift [5-9]. A recent experimental study [10] has demonstrated that mode splitting of WGM can be used as a highly accurate method for measuring single nanoparticle size, modeling from single particle perturbation [11]. However, due to the rotational symmetry of circular microcavities, these techniques are not sensitive to the exact particle or molecule position. And the applications are strongly limited by nanosized precise manipulation of the fiber or waveguide position [4-10]. Here, we demonstrate a new concept of biosensing in high Q deformed microcavities. Results from our numerical simulations show that a deformed microcavity can provide ultrahigh sensitivity to a particle position as well as a refractive index change. Moreover, due to the unique properties of directional emission, this deformed microcavity opens a new possibility to detect signals of a single particle or molecule in the far field.

When a nanoparticle is placed on the boundary of a microcavity, the degenerated clockwise (CW) and counter clockwise (CCW) modes are lifted and couple each other [12, 13]. Thus, we need to take into account the amplitude of the CW and CCW modes together to describe the scattering effect of the nanoparticle. In this case, the perturbation on resonant fields should be proportional to $f^2(r)$, where f(r) is the field distribution on the boundary [11]. In a circular cavity, due to the rotational symmetry, the field is uniformly distributed along the boundary. Thus, neither spectra shift nor mode splitting is sensitive to the particle position. On the other hand, in a deformed microcavity, the breaking of rotational symmetry of the cavity shape also breaks the uniform field distribution along the cavity boundary and therefore enables the position dependence of the resonant properties.

This can be easily understood with a simple ray model. Figure 1 (a) shows the total internal reflection confined WGM in a circular cavity. The Poincare surface of section (SOS) is illustrated in fig.1 (b), where S and Sin$\chi$ correspond to the arclength coordinate along the circumference and the light incident angle of the total internal reflection, respectively. The blue line, which can be simply analogue to the field distribution of resonant mode, is found to be invariable along the cavity boundary. However, in a deformed microcavity, the straight line is distorted to curves, broken curves, islands, and finally chaotic sea [14]. In an example of limacon cavity ($r = R\times(1+\varepsilon\cos\theta)$) with $\varepsilon=0.35$, where R is the overall size and $\varepsilon$ is the deformation parameter, the trajectory are focused on several bouncing points (points 1-4) only on the cavity boundary in fig.1 (c) and (d). Therefore, the final field distribution is focused around these points on the cavity boundary, allowing the sensitivity to the scatter's position. In addition, the directional emission from this deformed microcavity enables the potential far-field detection [14-21].

We systematically tested our concept using a finite element method (COMSOL Multiphysics 3.5a). We kept the limacon shape because ultrahigh Q factors and directional emission have been proposed [17] and experimentally observed [18-21]. For simplicity, we reduced our system to a two-dimensional structure and the polarization is limited to TE (E in plane). The refractive index of microcavity is 3.13, which is similar to our effective refractive index of GaAs. [18]

We first calculated the eigenvalues and observe several types of high Q modes as shown in Fig. 2(a). The red circles represent rectangle series of high Q modes, which are confined by wave localization [22] on the rectangle unstable periodic orbits (UPO) [14]. The Q factor obtained from such modes is greater than $10^8$. The corresponding field distribution is shown in Fig. 2(b), where the black solid lines show the UPO. Another type of high Q modes are marked as the green squares in fig.2 (a). Their Q factors are slightly lower than those of the rectangle orbit modes, but still higher than $10^6$. The corresponding field distribution and orbits are shown in fig.2 (c). Hereafter, this kind of mode is referred to as a diamond mode. These two types of the resonant modes have the highest Q factors and are most possible candidates to lase as active biosensors. [8]

To test the binding-site dependence of this deformed cavity, we placed a nanoparticle with a radius of 50 nm on the cavity boundary, as shown in the inset of fig.3 (a). First, we studied the rectangle mode, which is marked by the black arrow in fig.2 (a). The Q factor is greater than $2.8\times10^8$, which is as high as those of most circular cavities, promising an ultrahigh sensitivity. Figure 3 (a) shows the position dependence of mode splitting on the position of the scatter. As the limacon shape has mirror symmetry along $\theta = 0$, we only studied the S from 0 to 0.5. The mode splitting varies with the scatter position and two peaks can be observed. The similar behavior of the spectra shift is shown in fig.3 (b). Two modes shift to longer wavelengths together because of the increase in the refractive index or the radius [3]. Meanwhile, the wavelength shifts are also position dependent. Two maximum points can be observed at the same S positions in fig.3 (a). To confirm the relationship between the spectra shift and the mode splitting, we plotted $f^2(r)$ of the rectangle mode along the perimeter in fig.3 (c). We found that the behaviors of the mode splitting and the resonant wavelengths are qualitatively consistent with $f^2(r)$. Both the position dependence and peak positions are similar. The slight difference can be caused by the effective mode volume, cavity loss channel, and absolute amplitude of the field.

We further tested our concept with the mode splitting and wavelength shift of the resonant modes along the diamond orbit in fig.2 (d). Results from our numerical simulations show that both mode splitting (fig.4 (a)) and spectra shift (fig.4 (b)) of the diamond mode are also dependent on the position of the scatter, similarly to fig.3. The only difference is that their maximum points follow the maximum field distribution of the diamond orbit, where the maximum points are around θ ~ 0, 180, 90, and 270, respectively (fig.4 (c)).

The strong position dependence provides the information about the exact position of a single particle or molecule on a cavity. In other words, the same nanoparticle will generate different spectra shift or mode splitting when it binds to a different location on the cavity boundary. Meanwhile, the deformed microcavity has a unique property - directional emission. In an active microcavity, this property enables the detection in the far field. Fig.3 (d) shows the far field patterns of the highest Q resonant mode on the rectangle orbit. Similar to our previous report, an obvious unidirectional emission can be observed along θ = 0 [18]. As the emission light is confined in a narrow angle range, the slight near field perturbation can be easily detected in the far field without a further precise controlled fiber or waveguide coupler.

Moreover, in a circular microcavity, as the CW and CCW modes overlap in real space, the perturbation of the field distribution in the CW and CCW modes are the same. On the other hand, in a deformed microcavity, because the Goos-Hanchen shift must be considered in small cavities, the CW and CCW modes shift in the opposite direction and are not exactly overlapped in the real space, especially at the maximum field positions of bouncing points [23]. Then the perturbation of the scatter at one point is not equal for the CW and CCW modes. If the CW and CCW are not real degenerated, four eigenvalues should be obtained. However, the final eigenvalues of superposition are only two in our simulation. This directly proves that the CW and CCW modes are indeed degenerated modes.

In summary, we have demonstrated a new concept about a deformed microcavity sensor, which is not only sensitive to the refractive index perturbation, but also to the location of the scatter. Moreover, owning to the directional emission from a deformed microcavity, the slight perturbation by a single particle or molecule can also be easily detected in the far field. Therefore, our design opens an easier way to gather more information. Thus, our results can provide a new guideline for designing intelligent microcavities for biomolecule sensing. We believe that our proposed new concept will find its real applications in ultrahigh sensitive biosensors in the near future.

Acknowledgement:

This project was supported in part by the ITRAC internal pilot grant mechanism of the Indiana University Simon Cancer Center and a grant from Purdue Research Foundation.

Figure Caption:

Fig.1: (a) WGM trajectory in a circular microcavity. (b) Poincare SOS showing the trajectory (blue line) in phase space. (c) Diamond trajectory in a limacon shaped microcavity. (d) Poincare SOS for the trajectory (blue dots) in phase space, where chaotic sea can be observed.

Fig.2: (a) Eigenvalues in the limacon shaped microcavity. Two types of high Q factors resonant modes are marked with the red open circles and the green open squares. (b) and (c) show the corresponding mode distribution (Hz) of the two resonant modes marked by the arrows in (a).

Fig.3: (a) Mode splitting (b) resonant wavelengths and (c) square of the field distribution versus the position of the scatter for the rectangle resonant mode. Inset: Schematic picture of a limacon cavity based biosensor. The solid line in (b) is the wavelength position of a bare cavity. Strong position dependence can be observed. Sym and Anti-sym denote original symmetric and anti-symmetric modes. A single particle is attached on the boundary of the cavity. (d) is the corresponding far field pattern. Unidirectional emission can be observed.

Fig.4: (a) Mode splitting, (b) resonant wavelengths, and (c) square of the field distribution versus the position of the scatter for the diamond resonant mode.

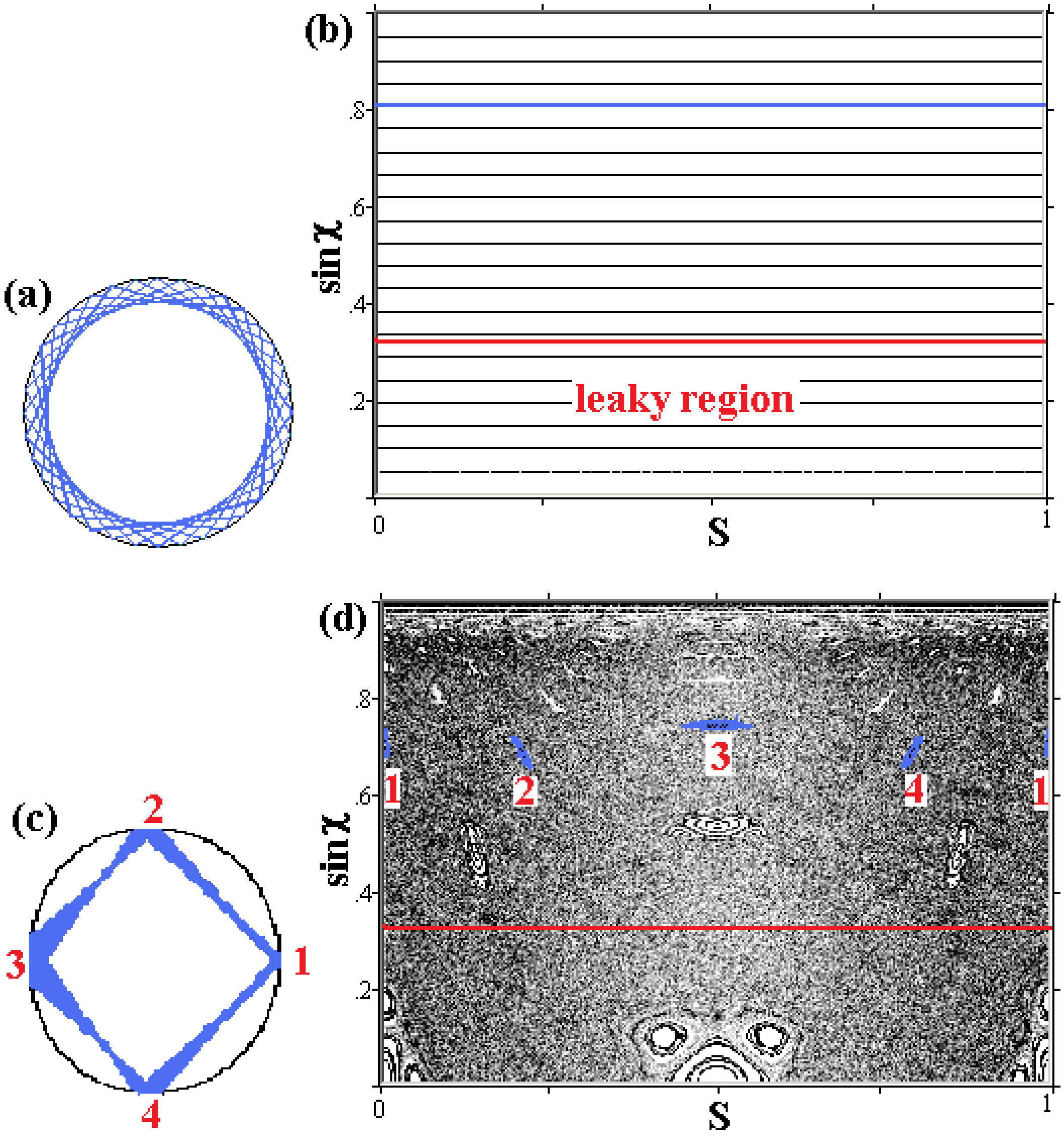

Fig. 1: Song et al

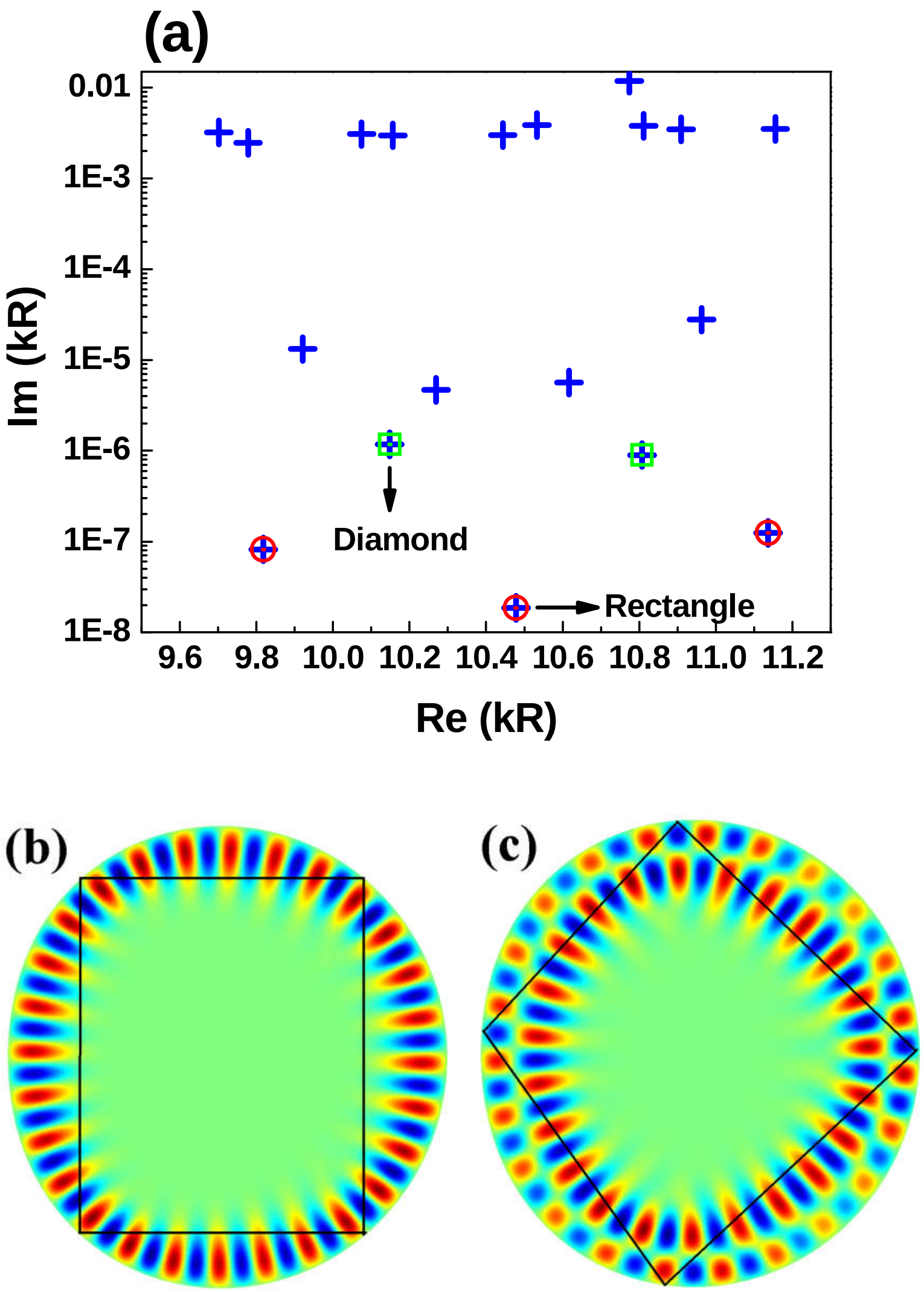

Fig.2 Song et al

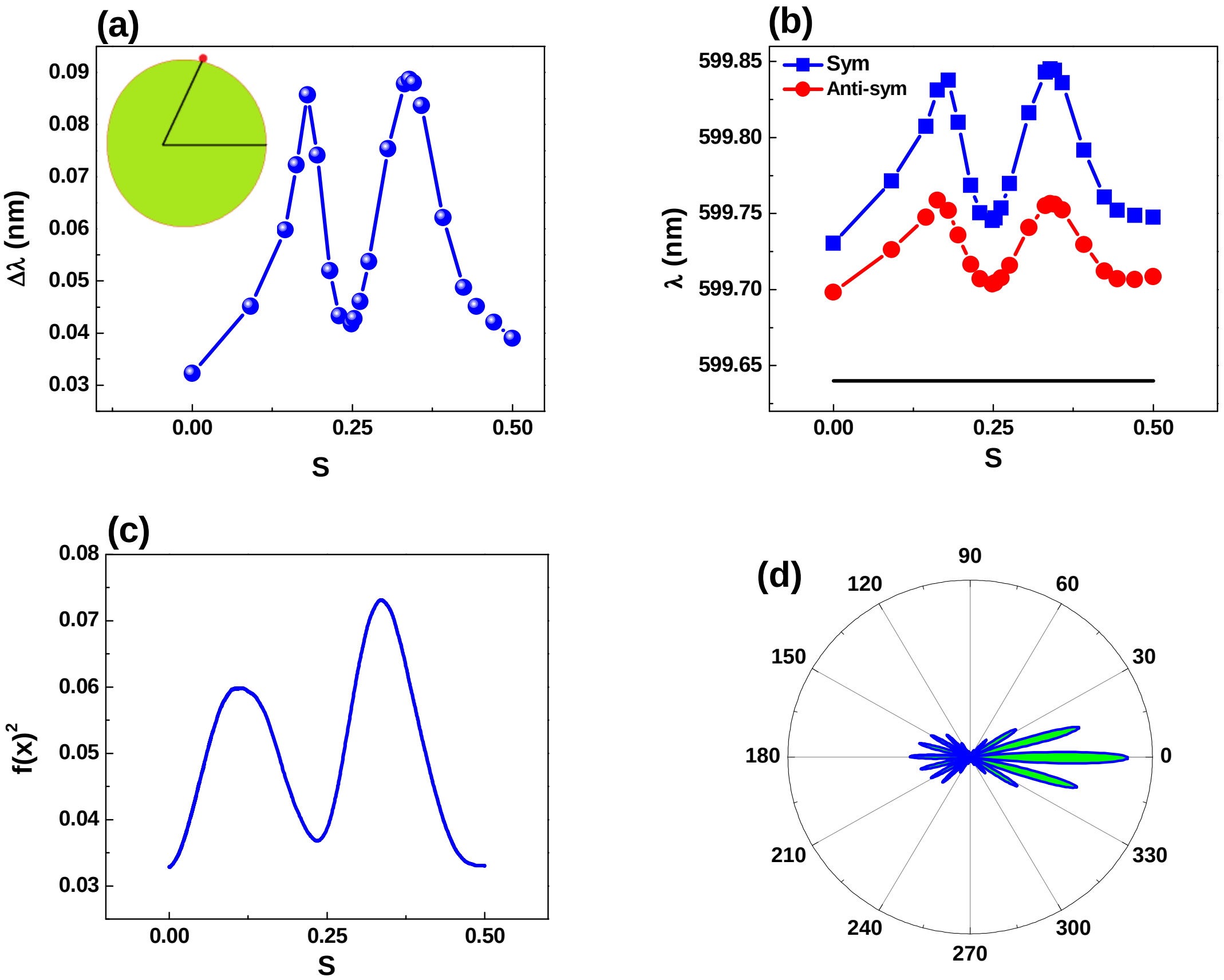

Fig.3 Song et al

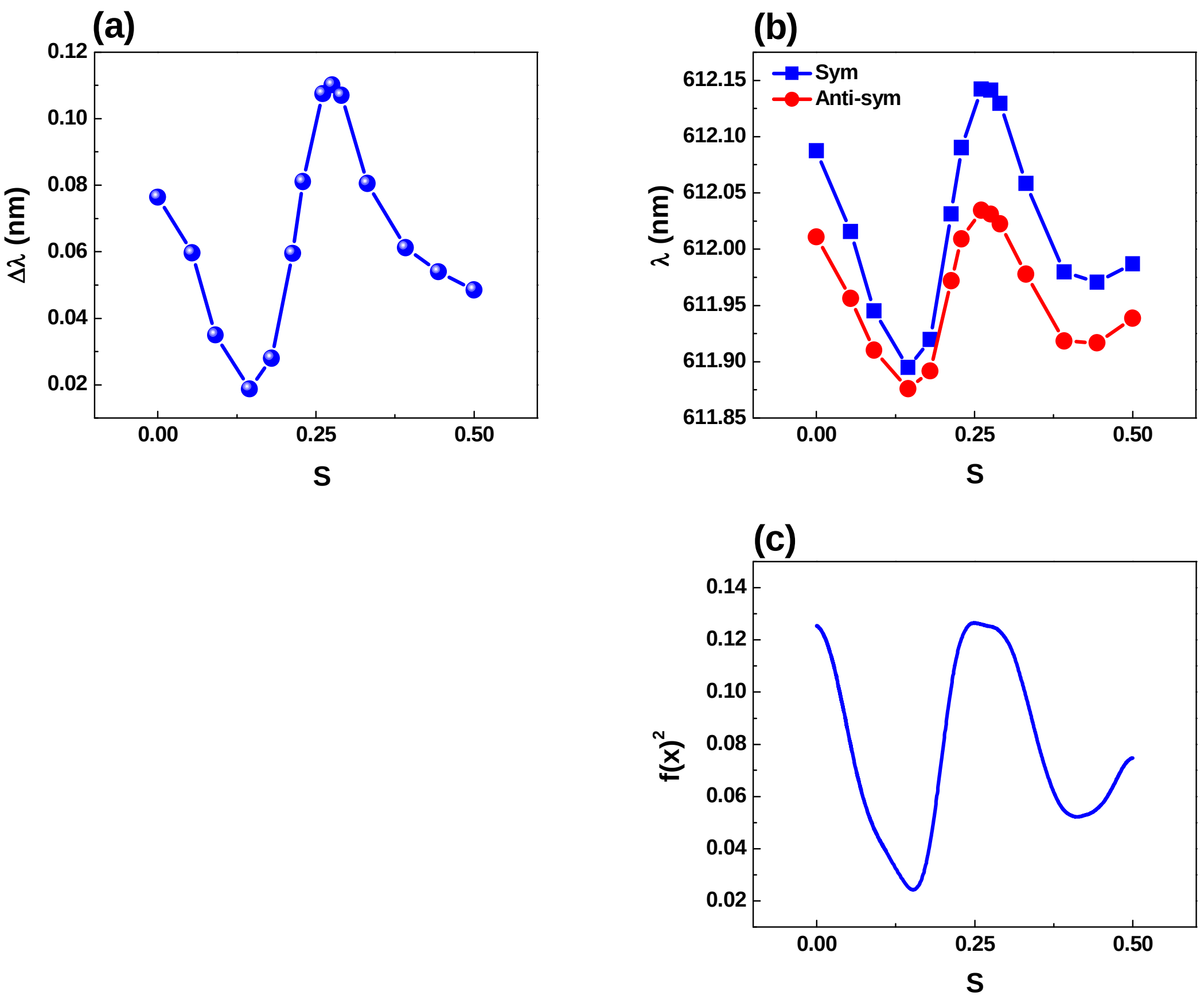

Fig.4 Song et al